\documentclass[final]{aipproc}

\usepackage{amsmath, amssymb}

\layoutstyle{8x11double}

\begin{document}

\title[Constraints on Hidden Photons]{New Constraints on Hidden Photons using Very High Energy Gamma-Rays from the Crab Nebula}


\author{Hannes-Sebastian Zechlin}{
  address={University of Hamburg, Institut f\"ur Experimental Physik, Luruper Chaussee 149, D-22761 Hamburg, Germany}
}

\author{Dieter Horns}{
  address={University of Hamburg, Institut f\"ur Experimental Physik, Luruper Chaussee 149, D-22761 Hamburg, Germany}
}

\author{Javier Redondo}{
  address={Deutsches Elektronen Synchrotron (DESY), Notkestrasse 85, D-22607 Hamburg, Germany}
}

\begin{abstract}
 Extensions of the standard model of particle physics, in particular those based on string theory, often predict a new $U(1)$ gauge symmetry in a hidden sector. The corresponding gauge boson, called hidden photon, naturally interacts with the ordinary photon via gauge kinetic mixing, leading to photon - hidden photon oscillations. In this framework, one expects photon disappearance as a function of the mass of the hidden photon and the mixing angle, loosely constrained from theory. Several experiments have been carried out or are planned to constrain the mass-mixing plane.
 
 In this contribution we derive new constraints on the hidden photon parameters, using very high energy $\gamma$-rays detected from the Crab Nebula, whose broad-band spectral characteristics are well understood. The very high energy $\gamma$-ray observations offer the possibility to provide bounds in a broad mass range at a previously unexplored energy and distance scale. Using existing data that were taken with several Cherenkov telescopes, we discuss our results in the context of current constraints and consider the possibilities of using astrophysical data to search for hidden photon signatures.
\end{abstract}

\maketitle


\section{Introduction}

 Typical extensions of the current standard model of particle physics often contain extra $U(1)$ gauge degrees of freedom. Many models based on string compactifications show that standard model particles are uncharged under the additional $U(1)$ symmetries. Therefore these degrees of freedom belong to a ``hidden sector'', i.e., an experimentally so far unobserved set of fields uncharged\footnote{In case of direct renormalizable couplings of the corresponding gauge boson to standard model matter precise measurements of the electroweak theory have shown that their masses must exceed a few hundred GeV \textit{\cite{Cavity:2008}}.} under the standard model gauge group. The existence of high mass particles charged under both the visible and hidden sectors (mediators) or gravity can produce nevertheless small interactions between the two sectors. Assuming that hidden sector particles are light\footnote{The masses typically considered belong to the sub-eV range.}, note that current accelerator based experiments can be largely insensitive to the subtle effects of their existence. Therefore in the recent past, many high-precision experiments have been dedicated to search for new light particles (e.g. \cite{Ahlers:2007rd} or \cite{DESY08} for an overview) and many ideas for new experiments are under consideration \cite{Patras:2008}.
 
 Natural models contain at least one hidden photon \cite{Holdom:1986} (or sometimes called paraphoton), i.e., the corresponding gauge boson to the new $U(1)_h$ gauge group. Here we consider a minimal theory with just \textit{one}\footnote{Of course theories with more than one additional $U(1)$ gauge symmetry exist, but they would be more cumbersome to handle \cite{Okun:1982}.} $U(1)_h$ gauge group (\cite{Okun:1982}, \cite{Holdom:1986}, \cite{Ahlers:2007rd}, \cite{Jaeckel:2008fi}), in addition to the normal electromagnetic gauge. The most general low energy Lagrangian allowed by the symmetries is
 \setlength \arraycolsep{0.1em}
\begin{eqnarray} \label{lagrangian}
\mathcal{L} & = & -\frac{1}{4} F^{\mu \nu} F_{\mu \nu} - \frac{1}{4} B^{\mu \nu} B_{\mu \nu} - \frac{\sin \chi}{2} F^{\mu \nu} B_{\mu \nu} + \nonumber \\
& & + \frac{\cos^2 \chi}{2} \mu^2 B^{\mu} B_{\mu}
\end{eqnarray}
 where $F_{\mu \nu}$ is the field strength tensor for the ordinary photon gauge field $A_{\mu}$, defined by \mbox{$F_{\mu \nu} = \partial_{\mu} A_{\nu} - \partial_{\nu} A_{\mu}$}, and $B_{\mu \nu}$ the same tensor for the hidden photon field $B_{\mu}$. The third term, also allowed by gauge invariance, corresponds to a non-diagonal kinetic term, the so-called kinetic mixing, where $\chi$ is the mixing angle between photons and hidden photons. We assume $\chi$ to be small. The last term describes a possible mass $\mu$ of the hidden photon. It can arise either from Higgs or St\"uckelberg mechanisms, but in the former case the model  suffers from additional constraints \cite{Ahlers:2008}.
 
 The non-zero kinetic mixing states that the $A_\mu$ and $B_\mu$ fields are non-orthogonal. The transformation to an orthogonal basis, with canonical kinetic term, is done by the redefinition
 \begin{equation}
 B_{\mu} \rightarrow S_{\mu} - \sin \chi A_{\mu}
 \end{equation}
 Moreover, one observes that \eqref{lagrangian} now contains a \mbox{non-diagonal} mass term that mixes photons with hidden photons (find the relevant part of the redefined $\mathcal{L}$ below, expanded for $\chi \ll 1$) 
 \begin{equation}
 \mathcal{L} = \dots + \frac{1}{2} \mu^2 \left( S^\mu S_\mu - 2 \chi S^\mu A_\mu + \chi^2 A^\mu A_\mu \right)
 \end{equation}
 leading to vacuum $\gamma$-$\gamma_s$ oscillations (if hidden photons are not massless). Here $\gamma_s$ is the quantum of the field $S_\mu$, that being orthogonal to the photon is completely sterile with respect to electromagnetic interactions. Note also that the diagonalization causes a multiplicative renormalization of the electric charge.
 
 It is interesting to note that the oscillation effect is completely analogous to the phenomenology of neutrino oscillations. Thus the oscillations of photons open the possibility to search for hidden photons via, e.g., ''light shining through a wall`` (LSW) experiments (\cite{Okun:1982}, \cite{Cavity:2008}, \cite{Ahlers:2007rd}). 
 
 In order to compute the oscillation probability at a distance $L$, one has to solve the equations of motion for the Lagragian to find the propagation eigenstates (see e.g. \cite{Ahlers:2007rd} for a short review). One obtains the following result for the oscillation probability (in natural units):
 \begin{equation} \label{oscprob}
 P_{\gamma \rightarrow \gamma_s}(L) = \sin^{2} (2\chi) \sin^{2} \left( \frac{\mu^2}{4E} L \right),
 \end{equation}
 where $E$ stands for the energy. Hence the oscillation length is observed to
 \begin{equation}
 L_{osc} = \frac{4 \pi E}{\mu^2} \simeq 8 \left( \frac{E}{\textnormal{TeV}} \right) \left( \frac{\mu}{10^{-7}\textnormal{eV}} \right)^{-2} \textnormal{kpc}.
 \end{equation}
 
 The values of the mixing parameters of hidden photons $\chi$ and $\mu$ are unspecified from theory and experimentally unknown because these particles have not been detected until today. However, there already exists a broad range of restrictions on these parameters provided by several experiments. The stronger constraints originate from tests of the Coulomb law, CMB measurements, LSW experiments, and searches of $\gamma_s$'s radiated from the Sun (see e.g. \cite{Ahlers:2008} and references therein). Find a composite plot of the current bounds in \cite{Ahlers:2008}, Fig. 1.
 
 Considering the experiments, it turns out that the bounds were obtained using energies up to $\mathcal{O}(100\, \textnormal{GeV})$ (LEP) and distances up to $\mathcal{O}(1\,\textnormal{AU})$ (solar searches), so it's worth mentioning that no constraints exist using the very high energy range ($\mathcal{O}(\textnormal{TeV})$) and distances with $\mathcal{O}(\textnormal{kpc})$. In this work we will provide such (astronomical) bounds using very high energy \mbox{(VHE, $E > 100\,\textnormal{GeV}$)} $\gamma$-rays from the Crab Nebula.
 
 The Crab Nebula (Messier 1) is a remnant of a supernova explosion that occurred in the year AD 1054. A pulsar (PSR B0531+21) is located near the geometrical center of the nebula. Today, the remnant is observed to be of the plerionic type at a distance of \mbox{$d_{\textnormal{c}} = (1.93 \pm 0.11)\, \textnormal{kpc}$} from Earth \cite{Trimble:1973}, with a bright continuum emission from radio to very high energy $\gamma$-rays, peaked in the near-infrared and optical range. The whole emission is predominantly produced by non-thermal processes, mainly by synchrotron and inverse Compton interactions of energetic electrons. The mechanism producing the VHE spectrum is inverse Compton scattering of accelerated electrons (up to PeV energies) on different low energy seed photon fields, dominated by the synchroton field (see e.g. Figure 10 in \cite{Aharonian:2004}).
 
 The broad-band VHE spectrum can be parametrised by (see \cite{Aharonian:2004} for the coefficients $p_i$)
 \begin{equation} \label{invCom}
  \log \left\{ \frac{\nu f_{\nu}}{\textnormal{erg }(\textnormal{cm}^2 \textnormal{ s})^{-1}} \right\} = \sum_{i=0}^{5} p_i \log^{i} \left( \frac{E}{\textnormal{TeV}} \right),
 \end{equation}
 where $\nu f_\nu = E^2 \frac{\mathrm{d}N}{\mathrm{d}E}$ is the differential energy-flux of the Crab nebula.
 
 Upper limits measured for the diameter of the nebula in the VHE regime are $\alpha_{\textnormal{c}} < 2'$ at \mbox{$E < 10\, \textnormal{TeV}$} and $\alpha_{\textnormal{c}} < 3'$ at $E > 10\,\textnormal{TeV}$ \cite{Aharonian:2004}. However, considering models of the VHE emission of the nebula (e.g. \cite{Aharonian:2004}) leads to diameters of $\mathcal{O}(\textnormal{arcsec})$, which we assume here. 
 
 Before we explain the method of giving new constraints, there remains one subtlety we have to mention briefly. The oscillation probability \eqref{oscprob} is calculated under the assumption that the photons can be represented as plane waves. Considering production and detection processes, this assumption does not hold under normal circumstances. Rather, \eqref{oscprob} must be calculated using quantum mechanical wave packets having a coherence length $D$. When working with wave packets of different masses $m_1 \neq m_2$, the velocities would differ by a factor \mbox{$\Delta \beta \approx \Delta m^2 / E^2$}, $\Delta m^2 = m_1^2 - m_2^2$, so they separate by $L \Delta \beta$ after traveling a distance L. Thus oscillations freeze out, if
 \begin{equation}
 L \geqslant L_{coh} = \frac{D}{\Delta \beta}
\end{equation}
(see \cite{Nussinov:1976}). In our case, $\Delta m^2 = \mu^2$. Detailed quantum mechanical calculations with wave packets reveal (see \cite{Giunti:1998}), that for relativistic particles
\begin{eqnarray}
 L_{coh} & = & \frac{4 \sqrt{2} \sigma_x E^2}{\mu^2} \\ 
 & \simeq & 1.8 \times 10^{22} \left( \frac{E}{\textnormal{TeV}} \right)^2 \left( \frac{\mu}{10^{-7} \textnormal{eV} }\right)^{-2} \left( \frac{\sigma_x}{\textnormal{m}} \right) \textnormal{pc}, \nonumber
\end{eqnarray}
where $\sigma_x \equiv \sqrt{\sigma_{xP}^2 + \sigma_{xD}^2}$, with $\sigma_{xP}, \sigma_{xD}$ the spatial uncertainties of the production and the detection process, respectively. To get oscillations, three conditions have to be satisfied: (a) $L_{coh} > d_{\textnormal{c}} \approx 2 \, \textnormal{kpc}$, (b) $L_{osc} < L_{coh}$, and (c) $\sigma_x \ll L_{osc}$.

For the production process one finds $\sigma_{xP}$ calculating the interaction length for electrons in the nebula, which interact via inverse Compton scattering with various seed photons. A lower limit is found assuming the Thomson regime, $\gamma \epsilon \ll m_e c^2$, with $\gamma$ the Lorentz-factor of the electron, $\epsilon$ the energy of the seed photon, $m_e$ the electron rest mass, and $c$ the speed of light in vacuum. In this case the interaction length is given by
\begin{equation}
 \lambda^{-1} = \sigma_{\textnormal{T}} \int_{\epsilon} \mathrm{d}\epsilon' n_{\textnormal{b}}(\epsilon'), 
\end{equation}
where $\sigma_\textnormal{T}$ is the Thomson cross section and $n_{\textnormal{b}}(\epsilon) = 
\frac{\mathrm{d}n}{\mathrm{d}\epsilon}$ is the differential (seed-) photon density. For a rough lower limit, approximating the main seed field (synchrotron) in \cite{Aharonian:2004} and using the extension of the nebula it reveals $\lambda \gtrsim 1 \,\textnormal{kpc}$.

Comparing $\lambda$ to the radius of the nebula in the VHE regime $r_{\textnormal{c}} \ll 1 \,\textnormal{pc}$, it should be clear that $r_{\textnormal{c}}$ is the right estimation for $\sigma_{xP}$, because $r_{\textnormal{c}} \ll \lambda$.

The spatial uncertainty of the detection process, $\sigma_{xD}$, does not contribute. Considering the detection process in detail one finds that $\sigma_{xD} \ll \sigma_{xP}$ (Zechlin, et al., in prep.).

Taking these results into account, one observes that the conditions given above hold for masses $\mu \lesssim 10^{-5} \,\textnormal{eV}$.

\section{Method}

As mentioned above, this work investigates the very high energy data available from the Crab nebula. The VHE regime is mainly covered by ground-based detection techniques, due to the fact that non-thermal \mbox{$\gamma$-ray} sources typically produce power-law type spectra in which the flux drops with increasing energy. As a consequence large effective detection areas of $\mathcal{O}(10^{5}\,\textnormal{m}^2)$ are required, sufficient to compensate for small photon fluxes. The data used here were taken with (stereoscopic) imaging air Cherenkov telescopes (IACTs), especially by HEGRA \cite{Aharonian:2004}, H.E.S.S. \cite{HessICRC:2007}, MAGIC \cite{Albert:2008}, and the Whipple 10\,m telescope \cite{Grube:2007}.

\begin{figure}[t!]
 \resizebox{.48\textwidth}{!}
 {\includegraphics{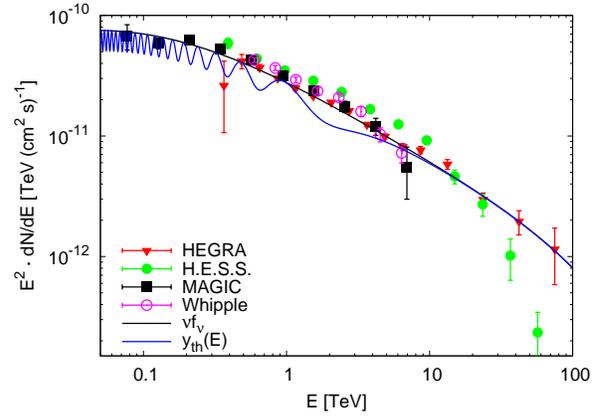}}
 \caption{The spectral data of four experiments is shown (references in the text). In addition, the signature of hidden photons $y_{\textnormal{th}} = (\nu f_{\nu}) \cdot (1-P_{\gamma \rightarrow \gamma_s}(d_{\textnormal{c}}))$ for specific parameters is demonstrated in comparison with the inverse Compton flux \mbox{$\nu f_{\nu}$ \eqref{invCom}.} Note that the data is not scaled (for further details consider the following section).} \label{fig:sig}
\end{figure}

If hidden photons of mass $\mu$ exist, the energy dependent oscillation probability \eqref{oscprob} influences the observable spectrum. Some photons will convert to hidden photons during their propagation to Earth which are not detectable because they initialize no air-shower. Due to the broad-band VHE spectrum that shows no intrinsic absorption lines in this region these spectral signatures should be measurable or could be used to constrain $\chi$ and $\mu$. The data and the expected spectral signature are shown in Fig.\,\ref{fig:sig}.

The differential spectra are measured as functions of energy-bins (see e.g. \cite{Aharonian:2004}), where $\overline{E}$ denotes the geometric mean energy of a bin with width $\Delta E$. Thus, one has to average \eqref{oscprob} (at the Crab distance $d_{\textnormal{c}}$) over the bin-size of the energy-bin centered on $\overline{E}$, giving
\begin{equation}
 \overline{P}_{\gamma \rightarrow \gamma_s}(\overline{E}, \Delta E) = \frac{1}{\Delta E} \int\limits_{\Delta E(\overline{E})} \mathrm{d}E \, P_{\gamma \rightarrow \gamma_s}(E, d_{\textnormal{c}})
\end{equation}
Therefore, the predicted energy-flux observed is
\begin{equation} \label{signature}
 \left. y(\overline{E}, \Delta E) = (\nu f_{\nu})\right\vert_{E=\overline{E}} \cdot (1-\overline{P}_{\gamma \rightarrow \gamma_s}(\overline{E}, \Delta E))
\end{equation}
 for the parameters $\chi$ and $\mu$, where the inverse Compton flux $\nu f_{\nu}$ from \eqref{invCom} was used.
 
Hence one has to fit the spectral signature \eqref{signature} to the data. This can be done applying a goodness-of-fit test, here the method of least squares is used (see e.g. \cite{Yao:2006}). In our case, $\chi^2_{\textnormal{lsq}}$ is given by the expression
\begin{equation} \label{chisq}
 \chi^2_{\textnormal{lsq}} = \sum\limits_{i=1}^N \left( \frac{y_i - y(E_i, \Delta E_i, \chi, \mu)}{\sigma_i} \right)^2
\end{equation}
where the sum runs over all data points, given by $(E_i,y_i,\sigma_i)$. $y_i$ stands for the measured energy-flux at energy $E_i$ with a statistical error $\sigma_i$. The method of least squares can be applied considering arbitrary confidence levels. The following results are calculated using the $68.3 \% $ confidence level which is constrained by \mbox{$\Delta \chi_{\textnormal{lsq}}^2 = 1$} (for one fit parameter).

Solving this numerically under the conditions explained above for every allowed mass $\mu$ ($\lesssim 10^{-5} \, \textnormal{eV}$, see above) one gets a fit value $\chi_{\textnormal{fit}}$ for $\chi$ for every $\mu$ ($\chi_{\textnormal{fit}}$ depends on the desired confidence level). Interpreting this in the disappearance approach it is clear that one can exclude all values $\chi \geq \chi_{\textnormal{fit}}$.

\section{Results}

Clearly, comparing the data of the different telescopes (see Fig.\,\ref{fig:sig}) it is worth mentioning that the results of the experiments differ among each other. Physically and within the experimental errors, all experiments must measure the same flux at a constant energy $E$. We choose to renormalize the energy scale of the instruments within their respective systematic uncertainties to avoid smearing of signatures. The scaling factor can be found by a fit of \eqref{invCom} on every data set.
\begin{ltxtable}[h]
\begin{center}
\caption{To avoid smearing of signatures the energy axis of every experiment has been rescaled by $E' = s \cdot E$. Values $\chi \geq \chi_{\textnormal{fit}}$ can be excluded. To check the goodness of the fit, the minimum reduced chi-squared is given for $\mu = 10^{-6} \,\textnormal{eV}$.}
\begin{tabular}{lccc}
\hline
\tablehead{1}{l}{b}{Telescope}
 & \tablehead{1}{c}{b}{Scalefactor s}
 & \tablehead{1}{c}{b}{$\chi_{\textnormal{fit}}$}
 & \tablehead{1}{c}{b}{$\chi^2_{\textnormal{red}}(n)$} \\
\hline
\hline
HEGRA & 1.000 & 0.0346 & 1.43(15)\\
H.E.S.S. & 0.921 & 0.0800 & 4.05(7)\\
MAGIC & 0.968 & 0.1165 & 0.82(9)\\
Whipple & 0.944 & 0.1036 & 1.21(7)\\
\hline
combined & - & 0.0343 & 1.60(41)\\
\hline
\end{tabular}
\label{tab:res}
\end{center}
\end{ltxtable}

For energies above $10\,\textnormal{TeV}$, the re-scaled spectra of H.E.S.S. and HEGRA still deviate from each other. Since the H.E.S.S. data are not consistent with the model considered (see \eqref{invCom}), we choose to ignore the data for energies above 10\,TeV.

\begin{figure}[b!]
 \resizebox{\linewidth}{!}
 {\includegraphics{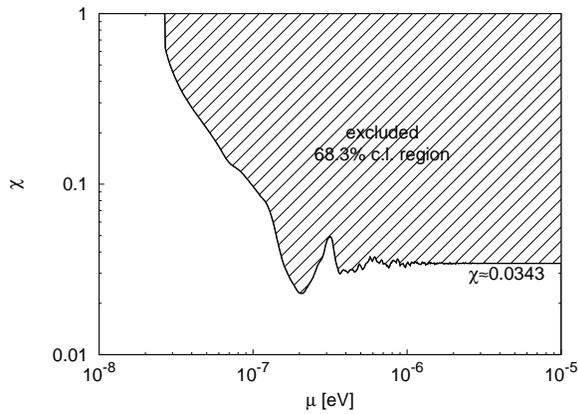}}
 \caption{Constraints on the mixing parameters of hidden photons using HEGRA, H.E.S.S., MAGIC, and Whipple data, 68.3\% C.L.. The marked parameter-region above the curve can be excluded.} \label{fig:res}
\end{figure}

Applying the method described above and using the HEGRA, H.E.S.S., MAGIC, and Whipple data we can conclude with a value $\chi_{\textnormal{fit}} \simeq 0.0343$ (the result converges to this value with increasing mass). The result is shown in Fig.\,\ref{fig:res}. Results for separate experiments can be found in \mbox{Tab. \ref{tab:res}}. To test the goodness of the fit, the minimum of the reduced chi-squared value, defined by \mbox{$ \chi^2_{\textnormal{red}}(n) := \chi^2_{\textnormal{lsq}}/n$}, is given in the table for a specific mass $\mu$, where $n$ is the number of degrees of freedom.

Due to experimental errors on the distance $d_{\textnormal{c}}$ and the energy $E_i$ of each data point, the error on the mass $\mu$ of every point excluded from the mass-mixing plane can be approximated to be 9\,\%.

\section{Conclusions}

Comparing our results to the constraints given in \cite{Ahlers:2008}, Fig.\,1, the limits obtained from measuring deviations from the Coulomb law are better. But the bounds given here are the best constraints on the hidden photon parameters using oscillation effects of photons directly. For the first time, we got astronomical limits considering new energy and distance ranges.


\begin{theacknowledgments}
We would like to acknowledge J.\,Grube for sending us the data, published in their proceedings (\cite{Grube:2007}). \mbox{H.-S. Zechlin} would like to acknowledge the ''Bundesministerium f\"ur Bildung und Forschung`` (BMBF) for making the participation on the conference possible.
\end{theacknowledgments}



\bibliographystyle{aipproc}   

\bibliography{biblio}

\begin{thebibliography}{16}
\expandafter\ifx\csname natexlab\endcsname\relax\def\natexlab#1{#1}\fi
\providecommand{\enquote}[1]{``#1''}
\expandafter\ifx\csname url\endcsname\relax
  \def\url#1{\texttt{#1}}\fi
\expandafter\ifx\csname urlprefix\endcsname\relax\def\urlprefix{URL }\fi
\providecommand{\eprint}[2][]{\url{#2}}

\bibitem[{Jaeckel} and {Ringwald}(2008)]{Cavity:2008}
J.~{Jaeckel}, and A.~{Ringwald}, \emph{Physics Letters B} \textbf{659},
  509--514 (2008).

\bibitem[Ahlers et~al.(2007)]{Ahlers:2007rd}
M.~Ahlers, H.~Gies, J.~Jaeckel, J.~Redondo, and A.~Ringwald, \emph{Phys. Rev.
  D} \textbf{76}, 115005 (2007).

\bibitem[DES(18-21 June 2008, http://axion-wimp.desy.de/)]{DESY08}
\emph{4th Patras Workshop on Axions, WIMPs and WISPs}, 18-21 June 2008,
  http://axion-wimp.desy.de/.

\bibitem[{Jaeckel}(2008)]{Patras:2008}
J.~{Jaeckel}  (2008), \eprint{arXiv:0807.5097}.

\bibitem[{Holdom}(1986)]{Holdom:1986}
B.~{Holdom}, \emph{Physics Letters B} \textbf{166}, 196--198 (1986).

\bibitem[{Okun}(1982)]{Okun:1982}
L.~B. {Okun}, \emph{Sov. Phys. JETP} \textbf{56}, 502--505 (1982).

\bibitem[Jaeckel et~al.(2008)]{Jaeckel:2008fi}
J.~Jaeckel, J.~Redondo, and A.~Ringwald, \emph{Phys. Rev. Lett.} \textbf{101},
  131801 (2008), \eprint{arXiv:0804.4157}.

\bibitem[{Ahlers} et~al.(2008)]{Ahlers:2008}
M.~{Ahlers}, J.~{Jaeckel}, J.~{Redondo}, and A.~{Ringwald}  (2008),
  \eprint{arXiv:0807.4143}.

\bibitem[{Trimble}(1973)]{Trimble:1973}
V.~{Trimble}, \emph{PASP} \textbf{85}, 579--585 (1973).

\bibitem[{Aharonian} et~al.(2004)]{Aharonian:2004}
F.~{Aharonian}, et~al., \emph{ApJ} \textbf{614}, 897--913 (2004).

\bibitem[{Nussinov}(1976)]{Nussinov:1976}
S.~{Nussinov}, \emph{Physics Letters B} \textbf{63}, 201--203 (1976).

\bibitem[{Giunti} and {Kim}(1998)]{Giunti:1998}
C.~{Giunti}, and C.~W. {Kim}, \emph{Phys. Rev. D} \textbf{58}, 017301 (1998).

\bibitem[{Kh\'{e}lifi} et~al.(2007)]{HessICRC:2007}
B.~{Kh\'{e}lifi}, et~al., \emph{H.E.S.S. ICRC 2007 proceedings} pp. 26--29
  (2007), \eprint{arXiv:0710.4057}.

\bibitem[{Albert} et~al.(2008)]{Albert:2008}
J.~{Albert}, et~al., \emph{ApJ} \textbf{674}, 1037--1055 (2008).

\bibitem[{VERITAS Collaboration: J.~Grube}(2007)]{Grube:2007}
{VERITAS Collaboration: J.~Grube}, \emph{ICRC 2007 proceedings}  (2007),
  \eprint{arXiv:0709.4300}.

\bibitem[{Yao} et~al.(2006)]{Yao:2006}
W.-M. {Yao}, et~al., \emph{Journal of Physics G} \textbf{33} (2006).

\end{thebibliography}

\IfFileExists{\jobname.bbl}{}
 {\typeout{}
  \typeout{******************************************}
  \typeout{** Please run "bibtex \jobname" to optain}
  \typeout{** the bibliography and then re-run LaTeX}
  \typeout{** twice to fix the references!}
  \typeout{******************************************}
  \typeout{}
 }

\end{document}